\documentclass[10pt,preprint2]{aastex}
\newcommand{\nuc}[2]{${}^{#2} \rm #1$}

\def\gtaprx {\lower .14ex\hbox{\rlap{\raise .9ex\hbox{\hskip .3ex
	{\ifmmode{\scriptscriptstyle >}\else
		{$\scriptscriptstyle >$}\fi}}}
	\kern -.4ex{\ifmmode{\scriptscriptstyle \sim}\else
		{$\scriptscriptstyle\sim$}\fi}}}
\def\ltaprx {\lower .14ex\hbox{\rlap{\raise .9ex\hbox{\hskip .3ex
	{\ifmmode{\scriptscriptstyle <}\else
		{$\scriptscriptstyle <$}\fi}}}
	\kern -.4ex{\ifmmode{\scriptscriptstyle \sim}\else
		{$\scriptscriptstyle\sim$}\fi}}}

\newcommand{\rr}{{\it r}}
\newcommand{\p}{{\it p}}

\newcommand{\s}{ \, {\rm s} }
\newcommand{\K}{ {\rm K} }

\newcommand{\km}{{ \rm km}}

\newcommand{\psec}{\,{\rm s}^{-1}}
\newcommand{\gpccm}{{ \rm g\,cm^{-3}}}

\newcommand{\del}[2]%
{\frac{\mathrm{d}{#2}}{\mathrm{d}{#1}}}
\newcommand{\Del}[2]%
{\frac{\mathrm{D}{#2}}{\mathrm{D}{#1}}}
\newcommand{\ddel}[2]%
{\frac{\mathrm{d}^2{#2}}{\mathrm{d}{#1}^2}}

\newcommand{\pdel}[2]%
{\frac{\partial{#2}}{\partial{#1}}}
\newcommand{\pddel}[2]%
{\frac{\partial^2{#2}}{\partial{#1}^2}}

\newcommand{\Ms}{M_{\odot}}
\newcommand{\DMs}{M_{\odot}\rm \,s^{-1}}

\shorttitle{Heavy element synthesis in a collapsar}
\shortauthors{Fujimoto et al.}

\begin{document}

\title{
Heavy element nucleosynthesis in a collapsar}

\author{
Shin-ichirou Fujimoto\altaffilmark{1},
Masa-aki Hashimoto\altaffilmark{2},
Kei Kotake\altaffilmark{3},
and Shoichi Yamada\altaffilmark{3,4}
}

\altaffiltext{1}{Department of Electronic Control, 
Kumamoto National College of Technology, Kumamoto 861-1102, Japan;
fujimoto@ec.knct.ac.jp}

\altaffiltext{2}{
Department of Physics, School of Sciences, 
Kyushu University, Fukuoka 810-8560, Japan}

\altaffiltext{3}{\textit{Science \& Engineering, Waseda University,
3-4-1 Okubo, Shinjuku, Tokyo 169-8555, Japan}}

\altaffiltext{4}{\textit{Advanced Research Institute for Science \&
Engineering, Waseda University, 3-4-1 Okubo,
Shinjuku, Tokyo 169-8555, Japan}}

%


\begin{abstract}
We investigate synthesis of heavy elements in a collapsar.
We have calculated detailed composition of magnetically driven jets
ejected from a collapsar, which is based on long-term, 
magneto-hydrodynamic simulations of a rapidly rotating massive star 
of 40$\Ms$ during core collapse.
We follow evolution of abundances of about 4000 nuclides
from the collapse phase to the ejection phase
through the jet generation phase 
with use of two large nuclear reaction networks. 
We find that the {\it r}-process successfully operates in the jets, 
so that U and Th are synthesized abundantly
when the progenitor has large magnetic field of $10^{12}$ G and
rapidly rotating core.
Abundance pattern inside the jets is similar compared to
that of {\it r}-elements in the solar system.
Heavy neutron-rich nuclei $\sim 0.01 \Ms$ 
can be ejected from the collapsar.
The detailed abundances depend on nuclear properties
of mass model, $\beta$-decay rate, and fission, 
for nuclei near the neutron drip line.
Furthermore, we find that {\it p}-nuclei are produced without seed nuclei:
not only light {\it p}-nuclei, such as \nuc{Se}{74}, \nuc{Kr}{78},
\nuc{Sr}{84}, and \nuc{Mo}{92}, but also 
heavy {\it p}-nuclei, \nuc{In}{113}, \nuc{Sn}{115}, and \nuc{La}{138}, 
can be abundantly synthesized in the jets.
The amounts of {\it p}-nuclei in the ejecta are much greater than those 
in core-collapse supernovae (SNe). In particular,
\nuc{Mo}{92}, \nuc{In}{113}, \nuc{Sn}{115}, and \nuc{La}{138}
deficient in the SNe, are significantly produced in the ejecta.
\end{abstract}




\section{Introduction}


A major fraction of elements heavier than iron are considered through 
a rapid neutron capture process ({\it r}-process) in an explosive event
\citep[e.g., ][and references therein]{qian03}.
Recent observations of metal poor stars (MPSs) have revealed that
{\it r}-elements have been produced in an early phase of metal enrichment
of the Galaxy~\citep[e.g., ][]{truran02}.
Abundance profiles of {\it r}-elements in some MPSs have similar patterns
of the {\it r}-elements in the solar system. 
The fact may suggest that
{\it r}-process can operate in an single site, or sites with 
a similar condition. 
It should be noted that 
{\it r}-elements are likely to produce in a site,
in which $\alpha$ elements as well as iron group elements
are not abundantly synthesized~\citep{qw03}.
Moreover, heavy {\it r}-elements (atomic number, $Z > 50$) 
suggest to have different origin of 
light {\it r}-elements ($Z > 40$)~\citep{wass96}.

Although various sites of {\it r}-process have been proposed, 
all the proposed scenarios have some deficiencies:
(1) Neutrino-driven winds, 
which are launched from a new-born neutron star
during core collapse of a massive star of 10-$25\Ms$, 
have been consider as a probable site of the {\it r}-process
~\citep[e.g., ][and references therein]{tera02}.
The scenario, however, requires an unlikely heavy and compact 
neutron star and 
the winds are possibly less abundant in heavy {\it r}-elements with 
mass number greater than 130~\citep{sumi00}.
(2) 
It is suggested that 
{\it r}-process operates inside ejecta from 
an O-Ne-Mg core, which is produced in a final stage of 
an 8-$10\Ms$ star and is expected to explode promptly 
during the core collapse~\citep{wheeler98}.
The {\it r}-process is shown to operate during the explosion of the star
if the explosion energy is an order of 
$10^{50}$-$10^{51} \rm ergs$~\citep{wanajo03}.
However, 
the star reveals to explode very weakly with an explosion energy 
of $2 \times 10^{49} \rm ergs$ and an ejected mass of $0.008\Ms$.
Moreover, 
the ejected mass of the {\it r}-elements is possibly too high 
for {\it r}-elements in the Galaxy. 
(3) A small iron core of a presupernova star of $\sim 11\Ms$ 
is expected to explode promptly~\citep[e.g., ][]{sumi01}.
Ejecta from the core have enough low electron fraction, or neutron-rich,
for {\it r}-process to operate successfully. 
The abundance profile of the ejecta can reproduce that of the solar system.
However, it is still debatable to explode the core.
(4) {\it r}-elements can be produced abundantly
in a neutron-rich ejecta from the merger of two neutron stars
~\citep[e.g., ][]{frt99}.
The composition is similar to that of the solar system.
This scenario is, however, inconsistent to the 
{\it r}-element abundances of MPSs in light of the Galactic 
chemical evolution
because of the rarity of the neutron star mergers~\citep{argast04}.
(5) 
Recently {\it r}-process calculations in MHD jets have been performed 
for a $13\Ms$\citep{nishimura05}.
Though the abundance distribution is well reproduced compared to the 
solar one, their results depend on the degree of neutronization 
that is very uncertain for multi-dimensional calculations.


On the other hands, any neutron capture processes 
cannot produce the {\it p}-nuclei that is 
35 neutron deficient stable nuclei with mass number $A \ge 74$.
The nuclei are considered to be synthesized by sequences of 
($\gamma, n$) photodisintegrations of {\it s}-nuclei ({\it s}-process seeds) 
processed during helium core burning in a massive star.
The scenario of the synthesis of the {\it p}-nuclei 
via the sequential photodisintegrations, or {\it p}-process, 
is proposed in the oxygen/neon layers 
of highly evolved massive stars 
during their presupernova phase~\citep{arnould76}
and during their supernova (SN) explosion~\citep{wh78}.
Extensive investigation of the {\it p}-process in core-collapse SNe
has shown that the overall abundance profile of the {\it p}-nuclei 
reproduces that in the solar system~\citep{rayet95}.
The scenario, however, has conspicuous shortcomings~\citep{rayet95};
(1) some {\it p}-nuclei, such as ${}^{92}\rm Mo$, ${}^{94}\rm Mo$, 
${}^{96}\rm Ru$, ${}^{98}\rm Ru$ and ${}^{138}\rm La$, 
are underproduced compared with those in the solar system
and (2) the ratio of {\it p}-nuclei to oxygen is smaller than 
that in the solar system.

During collapse of a massive star greater than 35-$40\Ms$,
stellar core is considered to promptly collapse to a black hole
~\citep{ww95,heger03}.
When the star has sufficiently high angular momentum before the
collapse, an accretion disk is formed around the hole and
jets are shown to be launched from the inner region of the disk
near the hole through magnetic processes~\citep{proga03,fujimoto06}.
Gamma-ray bursts (GRBs) are expected to be driven by the jets.
This scenario of GRBs is called a collapsar model~\citep{w93}.
Assisted by the accumulating observations implying the association between 
GRBs and the death of massive stars~\citep[e.g.,][]{galama98,hjorth03,zkh04}, 
the collapsar model seems most promising.
For accretion rates greater than $0.1\DMs$,
the accretion disk is so dense and hot 
that nuclear burning is expected to proceed efficiently.
In fact, an innermost region of the disk related to GRBs
becomes neutron-rich through electron capture
on nuclei~\citep{beloborodov03, PWH03, fujimoto03b, fujimoto04}.
Nucleosynthesis inside outflows from the neutron-rich disk 
has been investigated with steady, one-dimensional models of the disk 
and the outflows~\citep{PWH03, pruet04, fujimoto04, fujimoto05}.
Not only neutron-rich nuclei~\citep{fujimoto04} 
but {\itshape p}-nuclei~\citep{fujimoto05} are shown to be produced
inside the outflows.
However, abundances of the outflows are shown to highly depend on 
electron fractions of the outflows. 
The electron fractions are expected to change during 
the generation phase of the outflows near the base of the outflows.
Nevertheless, the electron fraction of the outflow at the base 
is assumed to be that of the disk at the base
~\citep{pruet04, fujimoto04, fujimoto05}.
Therefore, 
in order to evaluate change in the electron fractions of the outflows 
and to examine nucleosynthesis in a collapsar, 
we need non steady, multi-dimensional simulations of a collapsar from 
the collapsing phase of the star to the ejection phase of the outflows
from the star.



In the present paper, we propose jets from a collapsar, or 
rapidly rotating massive star during collapse to the black hole, 
as a {\itshape new r-process site}.
In order to investigate heavy element synthesis in a collapsar, 
we have calculated detailed composition of magnetically-driven jets
ejected from the collapsar, based on long-term, 
magnetohydrodynamic (MHD) simulations of a rapidly rotating, 
magnetized massive star of $40\Ms$ during core collapse, 
which has been performed in \citet{fujimoto06} recently.
We follow evolution of the electron fraction and abundances of 
about 4000 nuclides from the collapse phase to the ejection phase
through the jet generation phase, 
with the aid of two large nuclear reaction networks.

In \S 2 we briefly describe 
a numerical code for MHD calculation of the collapsing star, 
initial conditions of the star, and properties of jets from 
the star.
In \S 3, we firstly present
Lagrangian evolution of ejecta through the jets 
and then two large nuclear reaction networks, 
in which spontaneous and $\beta$-delayed fission is taken into account.
It is also shown in \S 3 that \rr-process operates inside the jets 
and that \p-nuclei that is deficient in core-collapse SNe 
can be produced inside the jets.
We discuss effects on {\it r}-process of our assumptions in a later expansion
phase of the ejecta and implications of collapsars
for the Galactic 
chemical evolution of {\it r}-elements and {\it p}-elements in \S 4. 
Finally, we summarize our results in \S 5.

\section{MHD Calculations of a Collapsar}\label{sec:mhd}

We have carried out two-dimensional, Newtonian MHD calculation of 
the collapse of a rotating massive star of 40$\Ms$, 
whose core is assumed to be collapsed to a black hole promptly.
We present our numerical models and results of the collapsar, 
in particular, the production and properties of jets.

\subsection{Input Physics and Numerical Code}\label{sec:model}

The numerical code for the MHD calculation
employed in this paper is based on the ZEUS-2D code~\citep{sn92}.
We have extended the code to include
a realistic EOS~\citep{kotake04} based on the 
relativistic mean field theory \citep{shen98}.
For lower density regime ($\rho < 10^5 \gpccm $), 
where no data is available in the EOS table with the Shen EOS, 
we use another EOS~\citep{bdn96}.
We consider neutrino cooling processes.
The total neutrino cooling rate is evaluated with 
a simplified neutrino transfer model based on 
the two-stream approximation~\citep{dpn02}, 
with which we can treat the optically thin and thick regimes 
on neutrino reaction approximately.
We ignore resistive heating, whose properties are highly uncertain,
not as in \citet{proga03}.
Spherical coordinates, ($r, \theta, \phi $) are used 
in our simulations and the computational domain is extended over
$50 \km \le r \le 10 000 \km$ and $0 \le \theta \le \pi/2$
and covered with $200(r) \times 24 (\theta)$ meshes.
Fluid is freely absorbed through the inner boundary of 50km, 
which mimics a surface of the black hole, whose mass is 
continuously increased by the mass of 
the infalling gas through the inner boundary.
We assume the fluid is axisymmetric and the mirror symmetry 
on the equatorial plane.
We mimic strong gravity around the black hole in terms of 
the pseudo-Newtonian potential~\citep{pw80}.

\subsection{Initial conditions}\label{sec:init-cond}

We set the initial profiles of the density, temperature and electron fraction 
to those of the spherical model of a 40$\Ms$ massive star 
before the collapse~\citep{hashimoto95}.
The radial and azimuthal velocities are set to be zero initially, 
and increase due to the collapse induced by the central hole and 
self-gravity of the star.
The computational domain is extended from the iron core to an inner
oxygen layer and encloses about 4$\Ms$ of the star.
The boundaries of the silicon layers between 
the iron core and the oxygen layers are located 
at about 1800 km (1.88$\Ms$) and 3900km (2.4$\Ms$), respectively.
We adopt an analytical form of the angular velocity 
$\Omega(r)$ of the star before the collapse:
\begin{equation}
 \Omega(r) = \Omega_0 \frac{R_0^2}{r^2 + R_0^2},
 \label{eq:omega0}
\end{equation}
as in the previous study 
of a collapsar~\citep{mizuno04a, mizuno04b} and SNe~\citep{kotake04}.
Here $\Omega_0$ and $R_0$ are parameters of our model.
Initial magnetic field, $B_0$, is assumed to be uniform, 
parallel to the rotational axis of the star.
We consider case with $\Omega_0 = 10\,\rm rad\psec$, 
$R_0$ = 1000km, and $B_0 = 10^{12}$G, or case with 
a strongly magnetized progenitor whose core rotates rapidly, 
because we pay attention to nucleosynthesis of heavy elements,
in particular {\it r}- and {\it p}-elements.
As shown in later (\S \ref{sec:abundance-particles}), 
{\it r}-process can operate successfully
and some {\it p}-elements be synthesized in jets from the star
during core collapse.

\subsection{Properties of Jets}\label{sec:jets}

We briefly describe results of our MHD calculation~\citep{fujimoto06}.
Lagrangian evolution of physical quantities inside jets, 
which is important for nucleosynthesis, 
is shown in \S 3.1, in detail.

We find that jets can be magnetically driven from the central region 
of the star along the rotational axis;
After material reaches to the black hole with high angular momentum
of $\sim 10^{17} \rm cm^2 \psec$, 
a disk is formed inside a surface of weak shock, 
which is appeared near the hole due to the centrifugal force
and propagates outward slowly.
The magnetic fields, which are dominated by the toroidal component, 
are chiefly amplified due to the wrapping of the field inside the disk
and propagate to the polar region along the inner boundary 
near the black hole through the Alfv{\'e}n wave.
Eventually, the jets can be driven by the tangled-up magnetic fields
at the polar region near the hole at $t = 0.20 \s$.

\section{Nucleosynthesis in a collapsar}\label{sec:nucleosynthesis}

We examine nucleosynthesis in a collapsar, based on the results
of our MHD simulations.
We firstly show Lagrangian evolution of ejecta through jets in detail
and then proceed nucleosynthesis inside the jets.

\subsection{Lagrangian evolution of ejecta through jets}\label{sec:ejecta}

In order to calculate chemical composition of material inside jets, 
we need Lagrangian evolution of physical quantities, such as 
density, temperature, and, velocity of the material.
We adopt a tracer particle method~\citep{nagataki97}
to calculate the Lagrangian evolution of the physical quantities
from the Eulerian evolution obtained our MHD calculations of a collapsar.
Particles are initially placed from the Fe core to an inner O-rich
layer. The numbers of the particles in a layer are weighted to the mass 
in the layer.
The total numbers of the particles are set to be 1000, with which 
we can follow ejecta through the jets appropriately.
We find that 59 particles can be ejected via the jets.

We have performed MHD calculation until $t = t_f$ ($t_f= 0.36 \s$).
After $t_f$, we assume that particles are adiabatic and 
expand spherically and freely.
Therefore, velocity, position, density, and temperature of a particle are
set to be $v(t) = v_0$, $r(t) = r_f +v(t)(t -t_f)$, 
$\rho(t) = \rho_f(r_f/r(t))^3$, $T(t) = T_f(r_f/r(t))$, respectively, 
where $v_0$ is constant in time and set to be $v_f$.
Here $v_f$, $r_f$, $\rho_f$, and $T_f$ are 
velocity, position, density, and temperature of each particle at $t_f$,
respectively.
We briefly discuss how the assumption on the expansion affects 
abundance change in the expansion phase of the ejecta, 
later in \S \ref{sec:expansion-phase}.

\begin{figure}[ht]
 \plotone{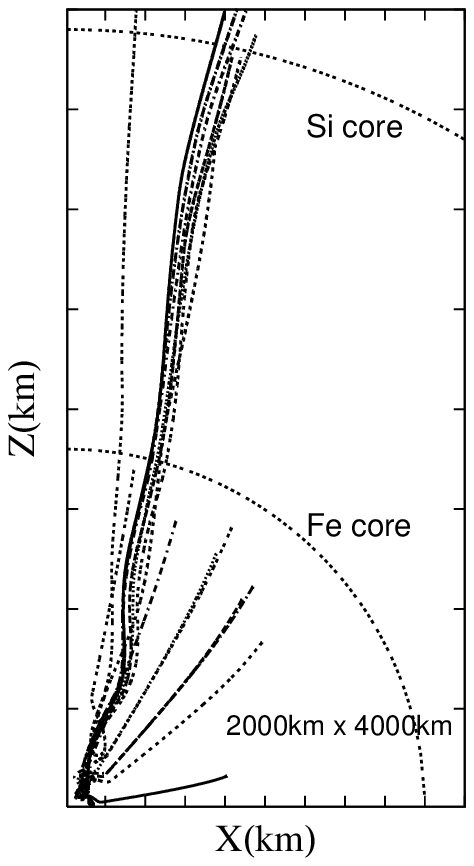}
 \plotone{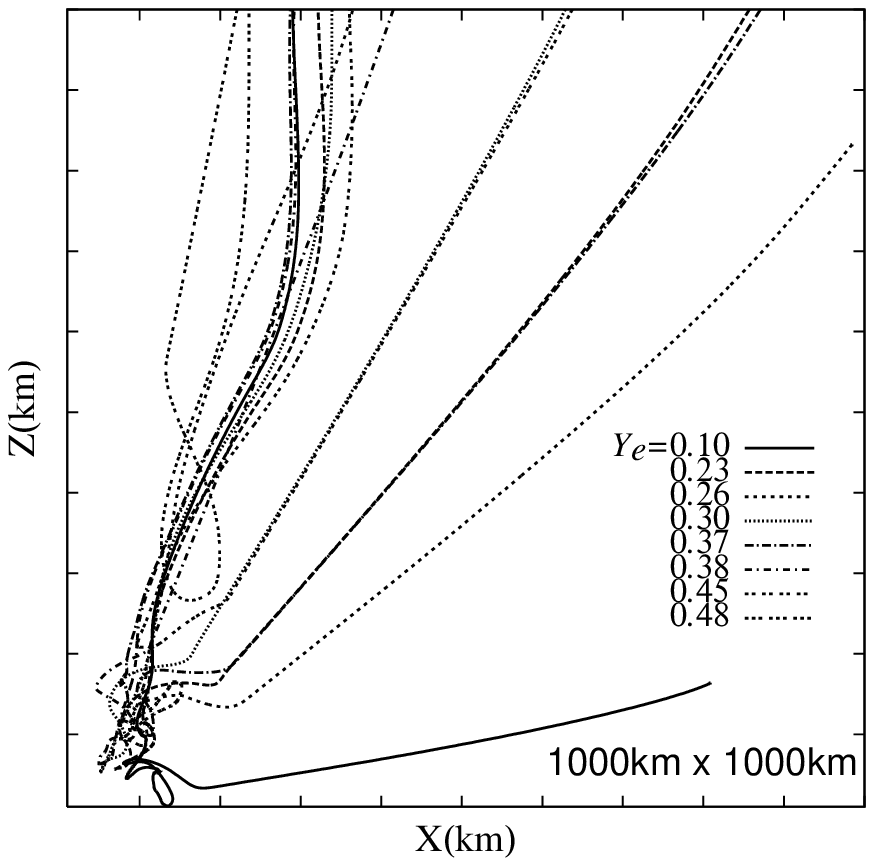}
\caption{
Trajectories of ejected particles.
We select the particles to have various values of $Y_e$.
The left and right panel cover the region of 
2000km $\times$ 4000km and 1000km $\times$ 1000km, respectively.
} \label{fig:trajectory}
\end{figure}

Figure \ref{fig:trajectory} shows trajectories of 
ejected particles with low $Y_e$,
where $Y_e$ is the electron fraction when the temperature 
of the particle is $9 \times 10^9$K.
We find that the particles with low $Y_e$ 
are initially located in the Fe core ($\le$ 1800km).
The trajectories of the particles are complex due to convective motion
near the equatorial plane~\citep{fujimoto06}.

\begin{figure}[ht]
 \plotone{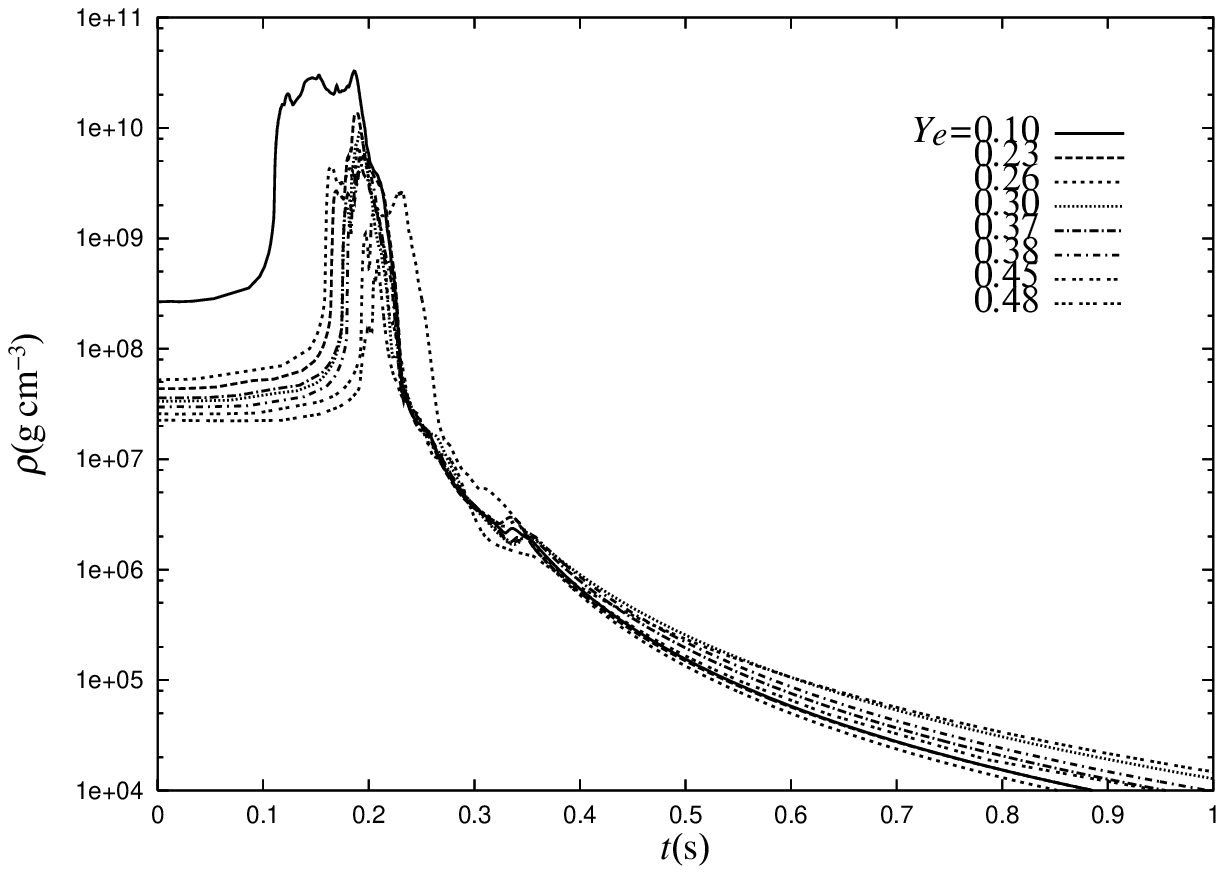}
 \plotone{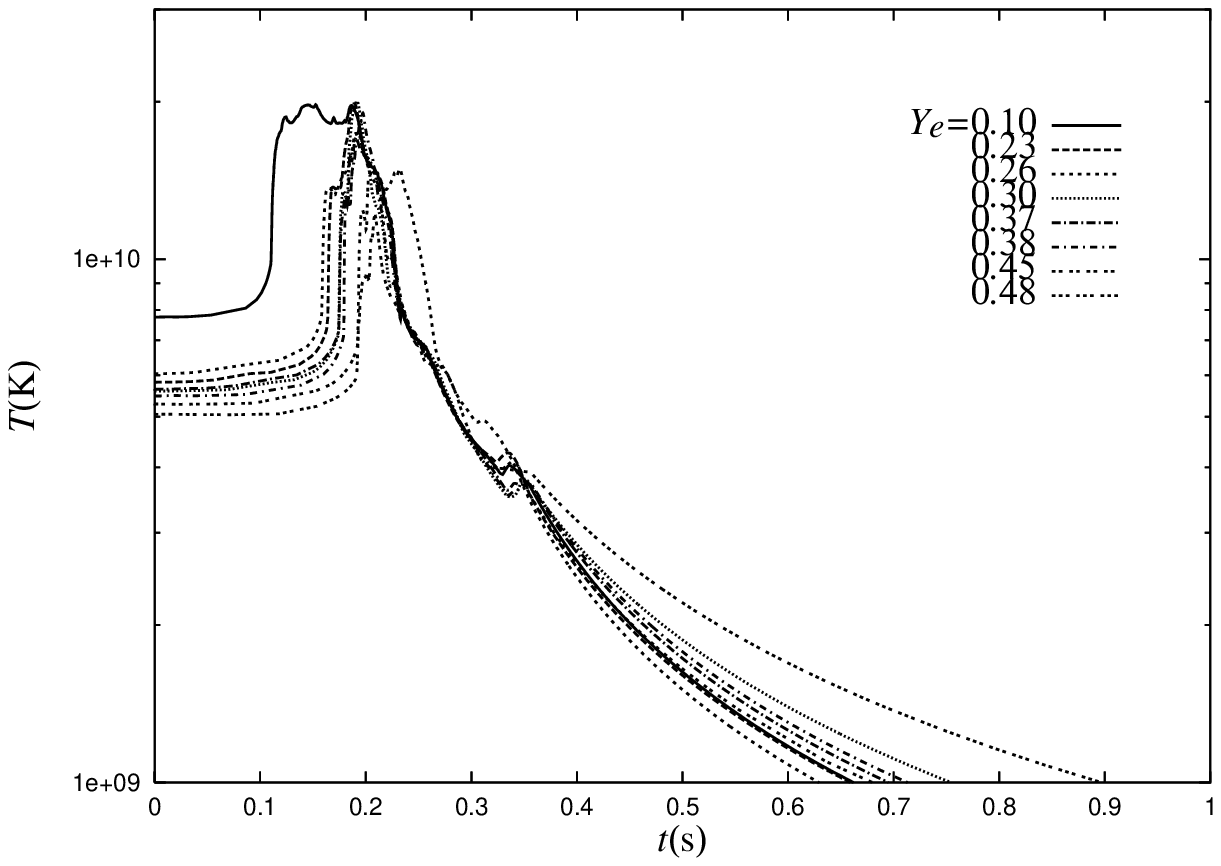}
\caption{
Evolution of density (top panel) and temperature (bottom panel)
for the particles shown in Figure \ref{fig:trajectory}.
} \label{fig:rho-T}
\end{figure}

Figure \ref{fig:rho-T} shows the evolution of 
density and temperature. 
Initial density and temperature of a particle is set to be
those of the presupernova in a layer where the particle 
is initially located.
As the particle falls near the hole, the density and temperature increase.
It should be emphasized that 
some particles become enough high densities ($> 10^{10} \gpccm$) and
high temperatures ($> 10^{10}$K) for protons to capture electrons, 
as we shall see later.
For the particle of $Y_e = 0.10$, 
the density and temperature stay at high levels during $t = 0.1$-$0.2\s$ 
because of convection motion near the black hole
(bottom panel in Figure \ref{fig:trajectory}).
The motion is important for decrease in the electron fraction, 
as we shall see later (Figure \ref{fig:ye}).

\subsection{Nuclear Reaction Network}\label{sec:network}

As shown in Figure \ref{fig:rho-T} (bottom panel), 
the ejecta can attain to high temperature greater than 
$9 \times 10^9 \K$ near the black hole, 
where material is in nuclear statistical equilibrium (NSE).
The abundances of the material in NSE
can be expressed with simple analytical expressions~\citep{clayton68,qian03}, 
which are specified by $\rho$, $T$, 
and the electron fraction as in \citet{fujimoto03b}.
The electron fraction changes through 
electron and positron capture on nuclei.
It should be emphasized that changes in the electron fraction 
through neutrino interactions can be ignored 
because the material is transparent for neutrinos
even in the most dense part of the computational domain~\citep{fujimoto06}.

In the relatively cool regime of $T < 9 \times 10^9 \K$, 
NSE breaks and the chemical composition 
is calculated with two large nuclear reaction networks, 
NETWORK A and NETWORK B, presented in \citet{nishimura05}.
The networks include about 4000 nuclei
from neutron and proton up to Fermium, whose atomic number, $Z$, 
is 100~\citep[see Table 1 in][]{nishimura05}.
The networks contain reactions such as 
two and three body reactions, various decay channels, and 
electron-positron capture.
The experimental masses and reaction rates are adopted if available.
Otherwise, 
theoretical nuclear data, such as nuclear masses, rates of two body reactions,
and $\beta$-decays 
are calculated with mass formula based on FRDM in NETWORK A 
but with that based in ETFSI in NETWORK B.
We note that theoretical rates of two body reactions related to nuclei 
with $Z \ge 93$ are same in the two network.

Spontaneous and $\beta$-delayed fission is taken into accounts
in the both networks.
Experimental half lives and branching ratios of the spontaneous fission 
are taken from \citet{JAERI} and \citet{Nudat2.0}. 
Theoretical formula of a half life \citep[eq. (23)]{KT1975}
with empirical fission barrier \citep{Mamdouh1999,Mamdouh2001}
is adopted for nuclei of $Z < 100$
without experimental evaluation of the half live.
We note that 
in \cite{nishimura05}, 
the half lives through the spontaneous fission are set to be $10^{-20} \s$
for all nuclei with the mass number, $A$, greater than 254.
Nuclei of $Z = 100$ decay through sequences of 
$\beta^-$ decays and/or fission.
However, in our networks, 
the nuclei decay only through fission 
because our networks contain nuclei up to $Z = 100$
and the $\beta^-$ channel of the nuclei of $Z = 100$ shut off artificially.
If we calculate the half life of the nuclei through the fission
with the theoritical formula of \citet{KT1975}, 
the nuclei, whose half life is long compared with the duration 
of postprocessing calculation, cannot entirely decay.
Therefore, for all nuclei of $Z = 100$, 
we tentatively set the half lives through the spontaneous fission 
to be $10^{-20}\s$.
We note that 
same branching ratios of $\beta$-delayed fission 
are taken from theoretical evaluation~\citep{Staudt1992} 
in the two networks although the theoretical ratios depend on 
an adopted mass formula~\citep{panov05}.
Empirical formula \citep[eq.(5)]{KT1975} is adopted about decay products
through the fission.
We note that the distribution of fission yields is asymmetric.

\subsection{Abundances of an ejected particle}\label{sec:abundance-particles}

Once we obtain the density and temperature evolution of a particle, 
we can follow abundance evolution of the particle
during the infall and ejection, 
through post-processing calculation 
using the nuclear reaction network presented in the previous subsection. 
The ejected particles are initially located in the iron core
(Figure \ref{fig:trajectory}).
As the particles infall near the black hole, 
the particles become hotter than $9 \times 10^9 \rm K$,
above which material in the particles is in NSE.
We therefore switch to the NSE code from the nuclear reaction
network to calculate abundance change of the particle. 
On the other hand, during the ejection, 
the temperature of the particles decreases and 
the particle becomes cooler than $9 \times 10^9 \rm K$.
We switch back to the reaction network from the NSE code
for the post-processing calculation of the particle composition.
We perform the post-processing calculation of nucleosynthesis
inside the jets until $\sim 10^{10}$ yr. 
Therefore, almost unstable nuclei has been decayed to their 
corresponding stable nuclei.

\begin{figure}[ht]
 \plotone{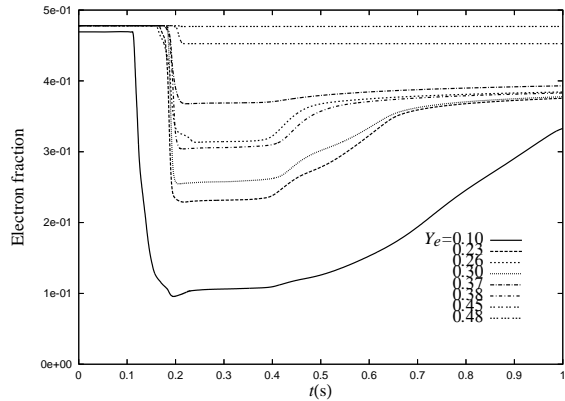}
\caption{
Evolution of the electron fraction
for particles shown in Figure \ref{fig:trajectory}.
} \label{fig:ye}
\end{figure} 

Figure \ref{fig:ye} shows evolution of the electron fraction 
for low $Y_e$ particles.
For $t < 0.2 \s$, 
the electron fractions decrease due to electron
capture on nuclei, mainly protons, near the black hole, 
where density and temperature of the particles are high.
On the other hands, 
for $t > 0.4 \s$, the electron fractions increase via 
$\beta^-$-decays of neutron-rich nuclei in particles
with $Y_e < 0.4$.
The fractions stay nearly constant values at $0.2 \s < t < 0.4 \s$, 
at which the particles are abundant in neutrons and long lived 
nuclei with the neutron number around 50. 
$\beta$-decay processes are therefore less important for 
decrease in the electron fraction at $0.2 \s < t < 0.4 \s$.
We find that there exist particles with $Y_e$ lower than 0.3.
The particles can be so neutron-rich that 
{\it r}-process operates efficiently.
Not only a region of high density and temperature 
but also staying in the region for a long time 
due to convective motion (Figure \ref{fig:trajectory}) 
are required for significant decrease in the electron fraction
of the particles.
The particle of low $Y_e \sim 0.1$ has low entropy per
baryon ($\sim $ 10-20 $k_{\rm B}$) where $k_{\rm B}$ is the Boltzmann
constant.
The condition is similar to that of ejecta from a massive star of 
$\sim 11\Ms$, in which the stellar material is expected to be 
exploded promptly during SN explosion~\citep{sumi01}.
The similar condition is also realized in a jet-like explosion from
an inner core of a rotating star of 13$\Ms$~\citep{nishimura05}.
The {\it r}-process can operate in the explosion~\citep{nishimura05}
but the minimum $Y_e$ in the explosion (= 0.158) 
is higher than that in the present study (= 0.10).
Therefore, U and Th cannot be synthesized abundantly
in the explosion, not as in the present study, 
as we shall see later (Figure \ref{fig:y-z-ETFSI}).

We have adopted weak interaction rates as those in \citet{ffn80,ffn82}.
However, recent shell-model calculations of the rates~\citep{langanke03} 
for nuclei in the mass range $A$ = 45-65 have resulted in 
substantial revisions to the old rate set of \citet{ffn80,ffn82}.
In order to estimate the effects of the revisions on $Y_e$ of particles, 
we have calculated evolution of the electron fraction of the $Y_e = 0.10$
particle using the revised weak interaction rates.
We find that 
the electron fraction calculated with the revised rates
is slightly larger than that with the old rates.
However, the difference in $Y_e$ with the two rate sets 
is only up to two percents.
This is because electrons are dominantly captured by protons
and the rates of electron capture on protons are comparable
between the two sets of the rates in high temperature and high 
density regimes.

\begin{figure}[ht]
 \plotone{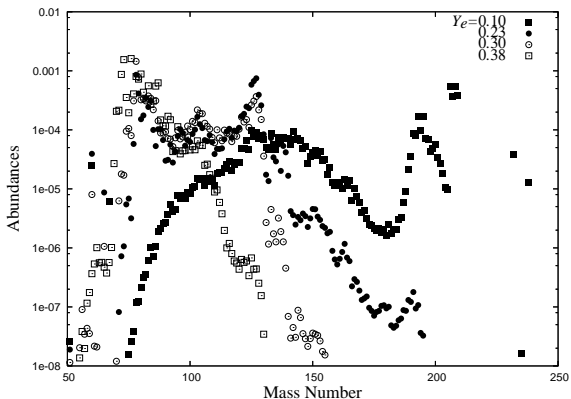}
\caption{
Abundances of nuclei with respect to the mass number $A$ 
for particles with low $Y_e$.
Filled squares, filled circles, open circles, and, open squares
represent the abundances of particles with $Y_e$ = 0.10, 0.23, 0.30,
and 0.38, respectively.} \label{fig:particle-abundance-ye}
\end{figure} 

\begin{figure}[ht]
 \plotone{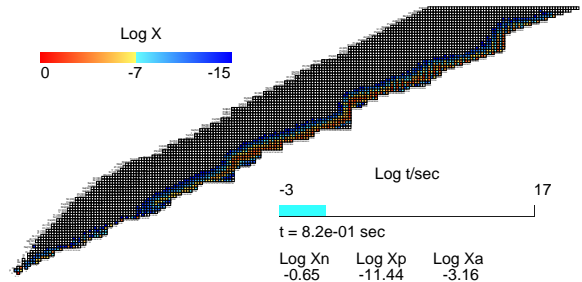}
\caption{
Mass fractions on the nuclear chart for a particle with $Y_e$ = 0.10
at $t = $ 0.82 s calculated with NETWORK B.} \label{fig:nuclear-chart}
\end{figure}

Figure \ref{fig:particle-abundance-ye}
shows abundances of nuclei with respect to $A$ 
for particles with $Y_e$ = 0.10, 0.23, 0.30, and 0.38 using NETWORK B. 
For a lower $Y_e$ particle, the particle becomes more neutron-rich
in the NSE phase, and the abundances of heavier nuclei increase.
We find that
the second-peak nuclei ($A \sim 130$) and 
the third-peak nuclei ($A \sim 195$) can be abundantly 
produced in particles with $Y_e$ less than 0.3 and 0.1, respectively.
We also find that
the particle with $Y_e$ = 0.10 is abundant in U and Th. 
The particle has enough low $Y_e$ for rapid neutron capture 
to proceed efficiently.
We find that the nuclear flow on the nuclear chart proceeds
near the neutron drip line and heavy neutron-rich nuclei with 
$Z \ge 95$ are abundantly synthesized in the particle 
(Figure \ref{fig:nuclear-chart}).
Therefore, nuclear properties of mass model, $\beta$-decay,
and fission, for nuclei near the drip line are important 
for abundance calculation of the $Y_e\sim 0.1$ particle.

\begin{figure}[ht]
 \plotone{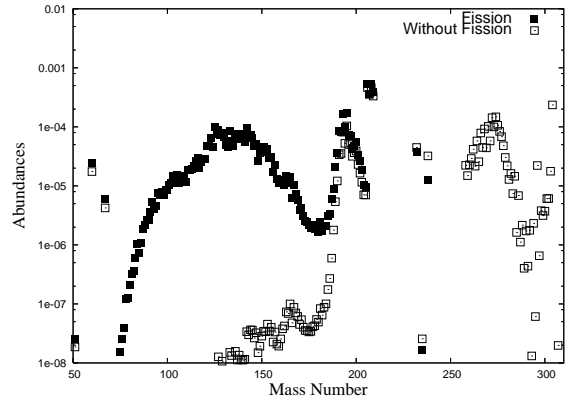}
\caption{
Abundances of the particle with $Y_e$ = 0.10 calculated with 
NETWORK B (filled squares) and a network same as NETWORK B 
but ignored all fission processes (open squares).
} \label{fig:particle-abundance-fission}
\end{figure} 
We have taken into accounts spontaneous and $\beta$-delayed fission 
in our nuclear reaction networks.
Fission is expected to have important role in the composition of
particles with low $Y_e$.
Figure \ref{fig:particle-abundance-fission} shows 
the abundances of the particle with $Y_e$ = 0.10 calculated with 
NETWORK B 
and a network same as in NETWORK B but ignored all fission processes.
We find that the abundances of nuclei with $A > 250$ as well as 
those with $60 < A < 180$ change significantly 
if the fission processes are ignored.
This is due to decay through fission and fission yields, 
whose mass numbers are abundantly distributed in $60 < A < 180$.
We note that fission recycling cannot take place in the particle.

\begin{figure}[ht]
 \plotone{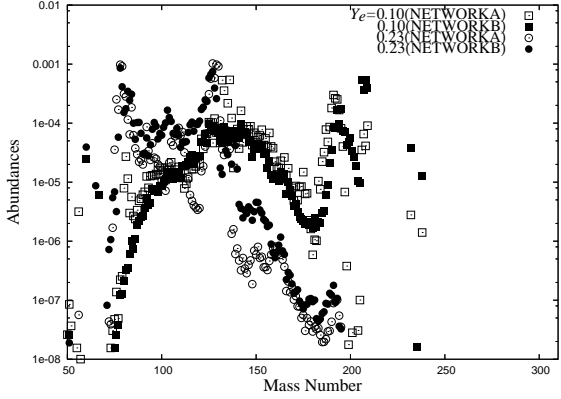}
\caption{
The abundances of particles with $Y_e$ = 0.10 and 0.23 calculated with 
NETWORK A and B presented in \S \ref{sec:network}.
Open squares, filled squares, open circles, and filled circles
represent abundances of the particle with $Y_e$ = 0.10 (NETWORK A), 
those with $Y_e$ = 0.10 (NETWORK B), 
those with $Y_e$ = 0.23 (NETWORK A), and 
those with $Y_e$ = 0.23 (NETWORK B), respectively.
} \label{fig:particle-abundance-mass}
\end{figure} 
Finally, in order to see dependence of particle composition 
on a nuclear mass model, 
we present the composition of particles with $Y_e$ = 0.10 and 0.23
using NETWORK A and B in Figure \ref{fig:particle-abundance-mass}.
We find that overall abundance profiles are similar for the two 
networks for the both particles.
Detailed profiles are, however, rather different for the two networks.
For the $Y_e=0.10$ particle, 
it is remarkable that heavy elements, such as Pb, Th, and U, with
NETWORK B are more abundant than NETWORK A.

\begin{figure}[ht]
 \plotone{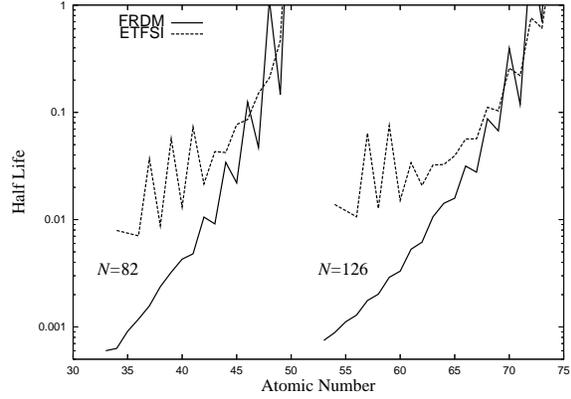}
\caption{
Half lives of neutron-rich nuclei with $N$ = 82 and 126.
The half lives are calculated with the two mass models
base on FRDM (solid lines) and ETFSI (dashed lines).
} \label{fig:half-lives}
\end{figure} 
The time-scale of a reaction sequence, which synthesizes heavy
neutron-rich nuclei greater than $A > 200$, mainly depends on 
half lives of the waiting-point nuclei of {\it r}-process, 
whose neutron number, $N$, is 82 or 126 and half live 
is long compared to the ejection time-scale of the particle.
The half lives based on FRDM are shorter than those with ETFSI
by an order of magnitude for nuclei with lower atomic number
(Figure \ref{fig:half-lives}), 
such as \nuc{Mo}{124} ($N = 82$) and \nuc{Gd}{190} ($N=126$), 
which are abundantly synthesized.
Hence, successive neutron capture can proceed to synthesize 
heavy neutron-rich nuclei with $A > 250$ in NETWORK A.
Most of nuclei with $A > 250$ decay to not U and Th 
but lighter nuclei via fission (Figure \ref{fig:particle-abundance-fission}).
Small fractions of the nuclei decay to U and Th through 
successive $\beta$-decays firstly followed by $\alpha$-decays.
On the other hand, 
neutron-rich nuclei with $A = $ 232, 235, and 238, 
can be abundantly produced in the ejecta with NETWORK B.
Most of the nuclei decay to U and Th through $\beta$-decays.
Small fractions of neutron-rich nuclei with $A > 250$ also decay 
to U and Th via a similar sequence of $\beta$ and
$\alpha$-decays as in case with NETWORK A.
This is because U and Th are underproduced in the ejecta
with NETWORK A compared to those with NETWORK B.

\subsection{Integrated abundances} \label{sec:abundance}

We shall move on mass-weighted abundances in the ejecta 
through the jets.

\begin{figure}[ht]
 \plotone{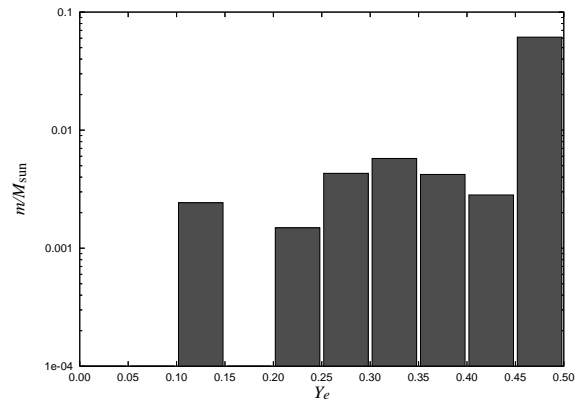}
\caption{
Mass distribution with respect to $Y_e$ in ejecta.
} \label{fig:mass-ye}
\end{figure} 
Figure \ref{fig:mass-ye} shows
mass distribution with respect to $Y_e$ in the ejecta.
The ejected mass is 0.081$\Ms$ through the jets.
We find that most particles have high $Y_e > 0.45$, 
whose mass is 0.061$\Ms$, and that
masses of particles in which third-peak nuclei can be produced
($Y_e \sim 0.1$) are 2.4$\times 10^{-3}\Ms$.
Masses of particles with $Y_e \le 0.4$ are 1.8$\times 10^{-2}\Ms$.

\begin{figure}[ht]
 \plotone{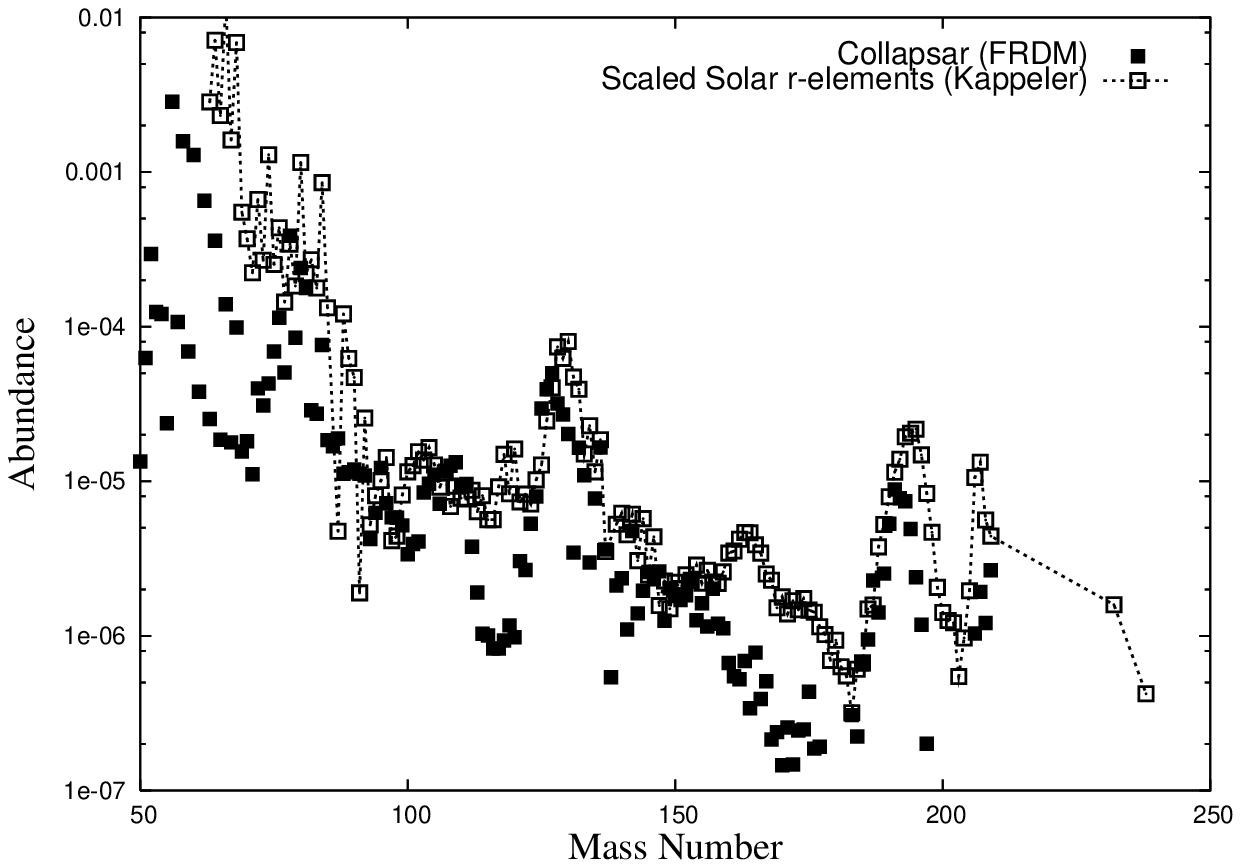}
\caption{
Integrated abundances in ejecta calculated with NETWORK A.
Filled squares and open squares with dotted line represent
abundances of the collapsar and \nuc{Eu}{153}-scaled abundances 
of solar {\it r}-elements~\citep{kap89}, respectively.
} \label{fig:y-a-FRDM}
\end{figure} 
Figure \ref{fig:y-a-FRDM} shows
mass weighted-abundances in the ejecta calculated with NETWORK A
with respect to the mass number.
The abundances of the solar {\it r}-elements~\citep{kap89}
are also presented in Figure \ref{fig:y-a-FRDM}.
We scale the abundances in the ejecta 
with the abundance of \nuc{Eu}{153}
because most Eu, up to 94.2\%, is produced via
{\it r}-process in the solar system~\citep{arlandini99}.
We find that the abundances in the jets well reproduce 
the solar pattern of {\it r}-elements.
It is clearly seen that nuclei around the second ($A\sim 130$) and 
third ($A\sim 195$) peaks can be synthesized in the ejecta via the jets.
However, nuclei with $160 < A < 180$ are underproduced 
compared with those in the solar system because of the lack of 
a $Y_e=$ 0.15-0.20 particle (Figure \ref{fig:mass-ye}).
Nuclei with $A < 70$ are also less abundant due to low $Y_e$ of 
the particles.

\begin{figure}[ht]
 \plotone{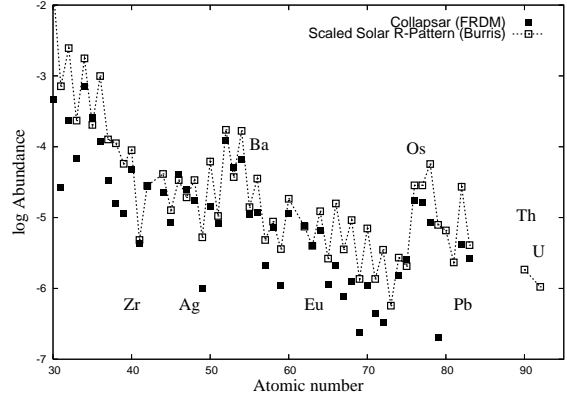}
\caption{
Integrated abundances in ejecta calculated with NETWORK A.
Filled squares and open squares with dotted line represent
abundances of the collapsar and Eu($Z=63$)-scaled abundances 
of solar {\it r}-elements~\citep{burris00}, respectively.
} \label{fig:y-z-FRDM}
\end{figure} 

Figure \ref{fig:y-z-FRDM} shows
mass weighted-abundances in the ejecta calculated with NETWORK A
with respect to $Z$ and the Eu ($Z=63$)-scaled abundances of 
the solar {\it r}-elements~\citep{burris00}.
An appreciable amount of {\it r}-elements of $Z \ge 40$ 
can be produced inside the jets.
The abundances of the jets well reproduce those in the solar system.
However, the detailed abundance profile of the ejecta are rather
different from that of the solar {\it r}-elements.
Elements with $65 \le Z \le 73$ and $79 \le Z \le 81$ are underproduced.
U and Th are also underproduced inside the jets.

\begin{figure}[ht]
 \plotone{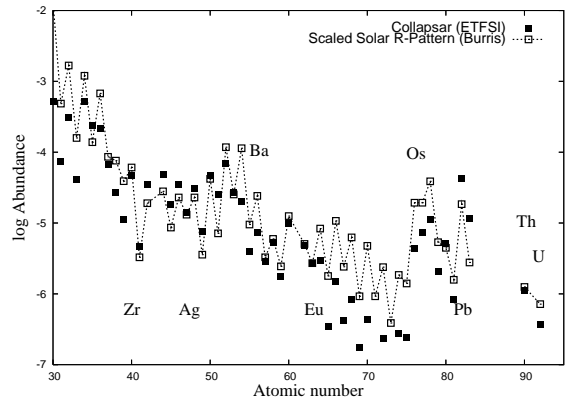}
\caption{
Same as Fig. \ref{fig:y-z-FRDM} but with NETWORK B.
} \label{fig:y-z-ETFSI}
\end{figure} 
If we use the another network, or NETWORK B, 
mass weighted-abundances in the ejecta slightly change 
but the overall profile is not largely different
from those with NETWORK A (Figure \ref{fig:y-z-ETFSI}).
The abundances of the jets well reproduce those in the solar system,
other than the ranges of $Z < 40$ and $70 \le Z \le 75$, 
in which the jet abundances are underproduced compared with those 
in the solar system.
It should be noted that U and Th, which are underproduced with NETWORK A, 
are comparably produced as the solar system.

\subsection{Abundances of nuclei other than {\it r}-process elements}
\label{sec:other-nuclei}

We present abundances of nuclei other than {\it r}-elements.
Note that our computational domain of MHD calculation 
extends over an oxygen rich layer ( $\le 10 000$km ).
Ejecta from an outer oxygen rich layer are not taken into account.
Ejected masses are therefore under-estimated
of elements lighter than iron, in particular 
$\alpha$-elements.

\subsubsection{$\alpha$- and iron group elements}

As shown in Figure \ref{fig:mass-ye}, 
most particles of $0.061 \Ms$ have high $Y_e > 0.45$.
Inside particles with $Y_e \sim 0.5$, 
$\alpha$-rich freezeout takes place as in the ejecta via a core-collapse SN.
\nuc{Ni}{56}, which decays to \nuc{Fe}{56}, is therefore abundantly produced
inside the ejecta.
We find that \nuc{Fe}{56} amounts to 0.013$\Ms$, 
which is comparable to masses of 
\nuc{Si}{28} (0.013$\Ms$), \nuc{S}{32} (9.4$\times10^{-3} \Ms$), 
\nuc{Ni}{58} (7.0$\times 10^{-3}\Ms$), and 
\nuc{Ni}{60} (6.4$\times 10^{-3}\Ms$).
Inside particles with $Y_e$ slightly lower than 0.5, or $\sim 0.49$, 
nuclei heavier than iron, such as \nuc{Ni}{62,64},
\nuc{Zn}{64,66,68}, and \nuc{Se}{78}, 
can be produced to be greater than $5 \times 10^{-4}\Ms$.

\subsubsection{{\itshape p}-elements}

We shall examine abundances and synthesis of {\it p}-nuclei 
in the ejecta through the jets.
It is suggested that {\it p}-nuclei are produced inside an accretion disk with 
high mass accretion rates, which can be realized during the collapse
of a rotating massive star~\citep{fujimoto03a}.
\begin{figure}[ht]
 \plotone{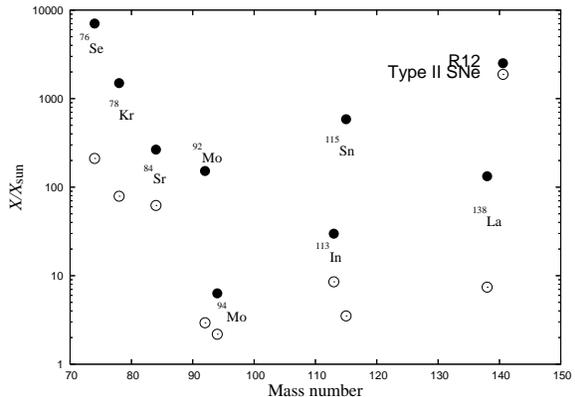}
\caption{
Mass fractions of {\it p}-nuclei abundantly produced in the ejecta through the jets.
The fractions are normalized by those in the solar system~\citep{anders89}.
IMF averaged abundances of {\it p}-nuclei from core-collapse SNe~\citep{rayet95} 
are shown with open circles.
} \label{fig:p-nuclei-mass-fraction}
\end{figure} 

We find that {\itshape p}-nuclei are 
significantly synthesized in the ejecta 
in spite of neutron-richness of the ejecta. 
Figure \ref{fig:p-nuclei-mass-fraction} shows 
the mass fractions of {\it p}-nuclei abundantly produced in the ejecta.
The abundances are calculated with NETWORK B. 
The fractions are normalized by those in the solar system~\citep{anders89}.
We also show mass fractions of {\it p}-nuclei from core-collapse SNe 
averaged over an initial mass function~\citep{rayet95}. 
We find that
not only light {\it p}-nuclei, such as \nuc{Se}{74}, \nuc{Kr}{78},
\nuc{Sr}{84}, and \nuc{Mo}{92}, but also heavy {\it p}-nuclei, 
\nuc{In}{113}, \nuc{Sn}{115}, and \nuc{La}{138}, 
are produced in the ejecta abundantly.
The abundances of {\it p}-nuclei in the ejecta are much greater than those in
core-collapse SNe~\citep{rayet95}.
It should be emphasized that 
\nuc{Mo}{92}, \nuc{In}{113}, \nuc{Sn}{115}, and \nuc{La}{138}, 
which are deficient in core-collapse SNe~\citep{rayet95}, 
are significantly produced in the ejecta.
We note that if we use NETWORK A, 
the abundances of heavy {\it p}-nuclei, 
\nuc{In}{113}, \nuc{Sn}{115}, and \nuc{La}{138}, 
decrease by a factor of 2, 8, and, 8, respectively.

\begin{figure}[ht]
 \plotone{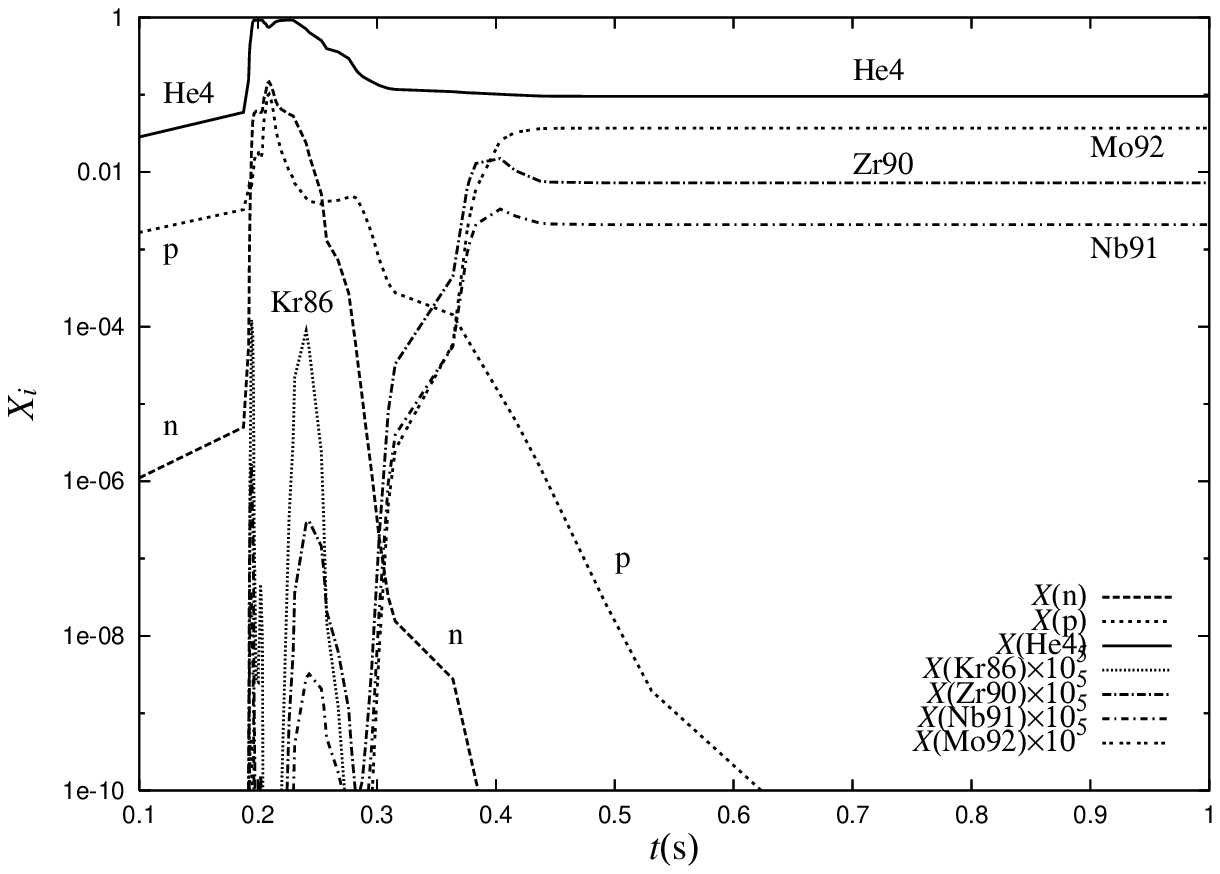}
\caption{
Abundance evolution of a tracer particle with $Y_e = 0.48$.
Mass fractions are shown of \nuc{Kr}{86}, \nuc{Zr}{90}, \nuc{Nb}{91}, 
and \nuc{Mo}{92}, whose abundances are multiplied by $10^5$.
Fractions are also presented of proton, neutron, and \nuc{He}{4}.
} \label{fig:p-nuclei-evolution}
\end{figure} 

In order to see production mechanism of the light {\it p}-nuclei, 
we present abundance evolution in a tracer particle with 
$Y_e = 0.48$ in Figures \ref{fig:p-nuclei-evolution}.
We note that the particle corresponds to the particle with $Y_e = 0.48$
in Figure \ref{fig:trajectory}-\ref{fig:ye}.
As the particle infalls near the black hole, the temperature 
raises up to $1.21\times 10^{10}$K (Figure \ref{fig:rho-T}).
Heavy nuclei are destroyed through photodisintegration
as the temperature increases.
Abundances are rapidly enhanced of proton, neutron, and \nuc{He}{4}.
As the temperature decreases during the ejection of the particle 
through the jets, 
heavy elements are re-assembled through neutron capture firstly, and then
proton and $\alpha$ capture.
We find that the light {\it p}-nuclei, \nuc{Se}{74}, \nuc{Kr}{78},
\nuc{Sr}{84}, and \nuc{Mo}{92} are produced via 
sequences of proton capture;
For \nuc{Mo}{92}, abundances are enhanced through a reaction sequence, 
\nuc{Kr}{86}($p$,$\gamma$)\nuc{Rb}{87}($p$,$\gamma$)\nuc{Sr}{88}($p$,$\gamma$)\nuc{Y}{89}($p$,$\gamma$)\nuc{Zr}{90}($p$,$\gamma$)\nuc{Nb}{91}($p$,$\gamma$)\nuc{Mo}{92}.
Note that \nuc{Kr}{86}, whose neutron number
is equal to the magic number of 50, 
are abundantly synthesized through neutron-capture 
because of neutron-richness of the particle ($Y_e$ = 0.48).
The reaction sequences, which produce the light {\it p}-nuclei, 
have been suggested to be realized in SNe~\citep{hmw91}.

On the other hands, 
the heavy {\it p}-nuclei, \nuc{In}{113}, \nuc{Sn}{115}, and \nuc{La}{138}, 
have different production mechanism as the light {\it p}-nuclei.
We find that the heavy {\it p}-nuclei can be produced only in a particle 
with $Y_e = 0.10$ and through fission processes.
In fact, the heavy {\it p}-nuclei cannot be synthesized 
if we ignore all decay channel through fission 
(Figure \ref{fig:particle-abundance-fission}).
\nuc{La}{138} are produced through fission directly but
\nuc{In}{113} and \nuc{Sn}{115} are $\beta$-decay products of 
fission yields, \nuc{Cd}{113} and \nuc{In}{115} 
(and their $\beta^-$-decay parents), respectively, 
whose half lives are $9.3 \times 10^{15}$ yr and $4.4 \times 10^{14}$yr.
We have performed the post-processing calculation of nucleosynthesis
inside the jets until $\sim 10^{10}$ yr.
If the calculation is performed for a much shorter time, say $\sim 10^{8}$ yr, 
the abundances are found to decrease by about two orders of magnitude.
It should be noted that we have adopted the empirical asymmetric 
distribution of fission yield~\citep{KT1975}, which can be fit 
experimental yields but its extrapolation into experimentally unknown 
region is questionable~\citep{KT1975}.
The abundances of the heavy {\it p}-nuclei are therefore highly uncertain.


\section{Discussion} \label{sec:discussion}

\subsection{Neutrino effects on nucleosynthesis}\label{sec:neutrino-effects}

We have ignored changes in the electron fraction 
through (anti-)neutrino capture on nuclei.
This is because the material is transparent for neutrinos
even in the most dense part of the computational domain, 
as shown in the previous study~\citep{fujimoto06}.
We have considered the collapse of a massive star to the black hole.
Neutrino emission is therefore found to be smaller than 
that in cases of core collapse to a neutron star~\citep{fujimoto06}
because of lower neutrino emission from the region near the black hole
compared with that near the neutron star.
Hence, jets from a collapsar are likely to be 
an appropriate site of the {\it r}-process because of 
low electron fraction and high neutron-richness 
due to low neutrino emission.


\subsection{Abundance change in an expansion phase}\label{sec:expansion-phase}

\begin{figure}[ht]
 \plotone{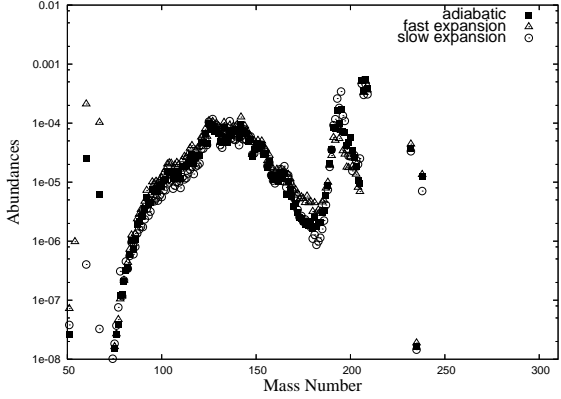}
\caption{
Abundances of the particle with $Y_e$ = 0.10 for various expansion 
velocities in the later phase.
Filled squares, open triangles, and, open circles represent
abundances of the particle for $v_0 = v_f$, $v_0 = 2 v_f$, and
$v_0 = 0.5 v_f$, respectively.
} \label{fig:particle-abundance-v0-ye0.1}
\end{figure}

As shown in Figure \ref{fig:nuclear-chart}, 
rapid neutron capture takes place during a later phase, or $t > 0.5 \s$,
which is greater than $t_f = 0.36 \s$.
Evolution of physical quantities in the phase is described in terms of
that of adiabatic free expansion.
In order to estimate dependence of particle composition on 
the assumption at the phase, we calculate 
abundances of a particle with $Y_e$ = 0.10 for various expansion 
velocities in the later phase (\S \ref{sec:ejecta}).
The kinetic energy of the jets
is larger than the internal and magnetic energies (Table 1).
Therefore, the velocity of the jets is unlikely to change appreciably
if the jets are adiabatic.
Figure \ref{fig:particle-abundance-v0-ye0.1} shows 
the abundances for cases with $v_0 = v_f$, $v_0 = 2 v_f$, and
$v_0 = 0.5 v_f$, which are represented with 
filled squares, open triangles, and, open circles, respectively.
For slower particles, neutron capture proceeds more efficiently.
Nuclei with $A \sim 170$ are less abundant but 
those with $A \sim 190$ are enhanced compared with those in 
a faster particle.
Nuclei heavier than U and Th are abundantly produced 
and eventually decay through fission.
However, the abundances are similar for three cases.
We conclude that the particle composition weakly depends 
on the assumption on the later phase expansion.
%


\subsection{Implications for Galactic chemical evolution}
\label{sec:gce}

\subsubsection{{\itshape r}-elements}



\begin{figure}[ht]
\plotone{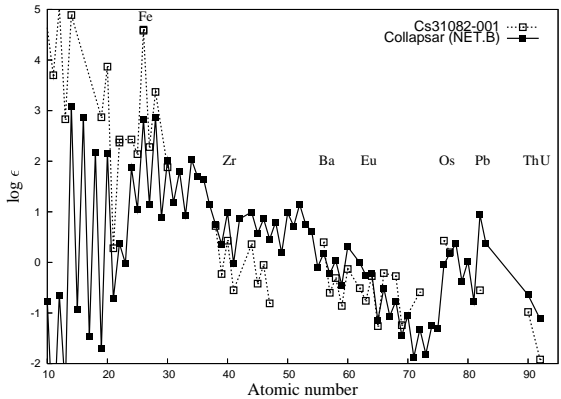}
\caption{
$\log \epsilon$ of the ejecta via the jets (open circles with dotted line)
and CS31082-001 (filled circles with solid line).
Note that $\log \epsilon$ are not scaled in the panel.
} \label{fig:eps}
\end{figure} 

Figure \ref{fig:eps} shows $\log \epsilon_i$~\footnote{
$\log \epsilon_{\rm A} = \log(N_{\rm A}/N_{\rm H})+12$ and
[A/B] = $\log(N_{\rm A}/N_{\rm B})- \log(N_{\rm A}/N_{\rm B})_{\odot}$
for elements A and B.}
for element $i$ of the ejecta through the jets, 
which are assumed to be uniformly mixed into inter stellar medium (ISM)
of mass, $M_{\rm ISM} = 4\times 10^5 \Ms$.
Note that $M_{\rm ISM} = 4\times 10^5 \Ms$ is slightly larger 
than mass of ISM polluted by an SN with the explosion energy of 
$10^{51} \rm ergs$~\citep{shigeyama98}.
Observed $\log \epsilon_i$ in an extremely metal poor star, 
CS31082-001 ([Fe/H]= -2.9), 
is also presented in Figure \ref{fig:eps}.
We find that $\log \epsilon_i$ of ISM mixed with the ejecta 
are similar to those in CS31082-001 for $Z \ge 38$.
However, for $Z < 30$, 
$\log \epsilon_i$ of the ISM are much less than 
those in CS31082-001.
Especially, iron is poor in the ISM mixed with the ejecta 
compared with that in CS31082-001 ([Fe/H] $\sim 5$ of the ISM).

\begin{figure}[ht]
\plotone{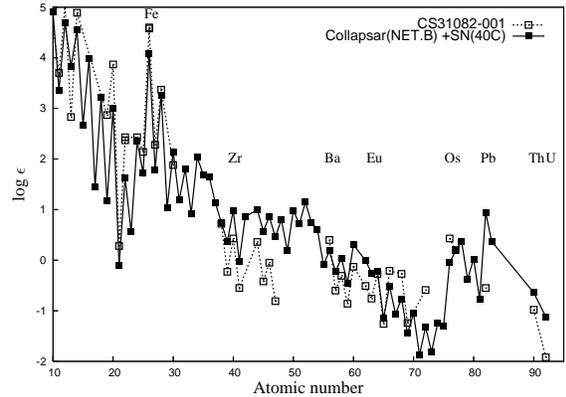}
\caption{
Same as Fig.~\ref{fig:eps} but for 
the ejecta via the jets mixed with ejecta via a hyparnova of $40\Ms$.
} \label{fig:eps-mps-col+sn}
\end{figure}

Moreover, we find that the observed $\log \epsilon_i$ of CS31082-001 
is well reproduced with those of the material in ISM 
for elements with $Z$ ranging from 10 to 92.
(Figure \ref{fig:eps-mps-col+sn}), if we assumed that 
the ejecta via the jets is uniformly mixed to the ISM of $M_{\rm ISM}$, 
which is pre-enriched by a hyparnova (HN) of $40\Ms$.
The ejected masses of nuclei from the HN are taken from model 40C 
in \citet{maeda03}, in which the star explodes aspherically with
extremely high explosion energy of $3.23 \times 10^{52}$ erg 
and ejects massive \nuc{Ni}{56} of 0.24 $\Ms$.
We note that 
nickel is over abundant if we adopt masses of SN ejecta of a star, 
which explodes spherically and can ejecta massive iron.

\subsubsection{{\itshape p}-elements}

The ejected mass from the collapsar is 0.081 $\Ms$.
The mass is smaller than 
ejected mass of {\it p}-nuclei, $M_{\rm PPL}$, of 0.15 $\Ms$
from a core-collapse SN of 13 $\Ms$~\citep{rayet95}.
However, the {\it p}-nuclei, 
\nuc{Se}{74}, \nuc{Kr}{78}, \nuc{Sr}{84}, 
\nuc{Mo}{92}, \nuc{In}{113}, \nuc{Sn}{115}, and \nuc{La}{138}, 
are produced much abundantly in the ejecta from the collapsar 
by 1-2 orders of magnitude compared with those in the
ejecta from the SN (Figure \ref{fig:p-nuclei-mass-fraction}).
The masses of the above {\it p}-nuclei from the collapsar, therefore, 
are comparable or greater than those from the SN.
The collapsar could be important for the Galactic chemical evolution
of the {\it p}-nuclei if the events are not rare compared with 
core-collapse SNe.

Moreover, 
{\it p}-process in oxygen-rich layers is also expected to operate 
in the collapsar, as in {\it p}-process layers (PPLs) in core-collapse SNe.
This is because {\it s}-process seeds of {\it p}-process exist
in the layers due to weak {\it s}-process during He core burning of the star
and the jets propagate in the layers to heat the layers to 
enough high temperatures for {\it p}-process to operate efficiently.
Therefore masses of {\it p}-nuclei can be enhanced via the {\it p}-process,
Ejected {\it p}-nuclei from the collapsar are the sum of those in the ejecta 
from oxygen-rich layers with the composition similar to 
that in PPLs in core-collapse SNe
and those in the jets from inner core, which abundantly 
contain {\it p}-nuclei deficient in the SNe.

Although ejected mass of {\it p}-nuclei is possibly smaller than $M_{\rm PPL}$
because of aspherical explosion in the collapsar, 
a ratio of the mass to ejected mass of oxygen can be much larger
than a ratio in the SNe.
This is because 
the jets from the inner core have small amounts of oxygen.
We note that the ratio in the SNe is problematically 
much smaller compared with that in the solar system~\citep{rayet95}.

\subsubsection{Event rates}

Finally, we briefly discuss on event rates of GRBs linked to the death 
of a massive star and jet-like explosion related to GRBs.
A current GRB rate is estimated to be $\sim 5 \times 10^{-5} \rm \, yr^{-1}$
in the Milky Way~\citep{firmani04}.
A GRB rate in an average galaxy is comparable to that in the Galaxy
and is much smaller than the rate of core-collapse SNe by a factor of 
several hundreds~\citep{podsiadlowski04}.
The GRB rate is however sensitive to the jet opening angle~\citep{podsiadlowski04}.
The rate rapidly increases at red shift $z \sim 1$ and 
is gradually enhanced at higher red shift~\citep{yonetoku04,murakami05}.
However, an estimated star formation rate
is constant at higher red shift, though the rate 
increases at $z \sim 1$ as the GRB rate~\citep{giavalisco04}.
Thus, the relative GRB rate to core-collapse SNe enlarges at the higher $z$.

Moreover, 
the jets in the current investigation are too slow ($\le 0.1c$) and 
heavy ($0.022\Ms$) to induce GRB.
Although this ``failed GRB'' is possibly hard to observe compared 
with normal GRBs, 
two failed GRBs are suggested to be associated with Type Ic SNe, 
SN2001em~\citep{granot04} and SN2002ap~\citep{totani03}.
A ratio of failed GRBs to Type Ic SNe is estimated to be 
$2/40 = $ 5\% for nearby 40 Type Ibc SNe~\citep{granot04}.
Note that the ratio can be much greater than the above estimate,
or $2/15 = $ 13.3\% if samples of Type Ibc SNe are limited to those 
with late-time ($> 100 \s$) observations, or 15 Type Ibc SNe~\citep{granot04}. 
Accordingly, we can estimate a failed GRB rate to 
$\sim 1.3 \times 10^{-4} \rm \, yr^{-1}$
with an estimate of a Type Ibc rate of
$1 \times 10^{-3} \rm \, yr^{-1}$~\citep{podsiadlowski04}.
Therefore, failed GRBs are possibly more frequent than normal GRBs.

In brief, the jets from the failed GRBs have much importance to 
chemical evolution of galaxies compared to normal GRBs because of 
greater ejection mass and frequency.
In the early universe, or $z > 1$, 
the jets could have a significant contribution to
chemical evolution of galaxies, in particular heavy elements, 
which cannot be abundantly synthesized in core-collapse SNe.

\section{Summary and Conclusions}

We have calculated detailed composition of magnetically-driven jets
ejected from a collapsar, based on long-term, 
magnetohydrodynamic simulations of a rapidly rotating massive star 
of 40$\Ms$ during core collapse.
The magnetic field of the star before the collapse
is set to be uniform and parallel to the rotational axis of the star.
We consider case with the magnetic field of $10^{12}$ G.
We follow evolution of abundances of about 4000 nuclei
from the collapse phase to the ejection phase
through the jet generation phase with the aid of a NSE code and 
two large nuclear reaction networks.
We summarize our conclusions as follows;
\begin{enumerate}
 \item The {\it r}-process successfully operates in the jets.
       Abundances of the jets have similar profile as those 
       of {\it r}-elements in the solar system, 
       though the detailed abundances are different.
       U and Th can be abundantly synthesized in the jets.
 \item Massive neutron-rich nuclei $\sim 0.01 \Ms$ 
       can be ejected from the collapsar.
 \item The abundances highly depend on nuclear properties of
       mass model, $\beta$-decay and fission, 
       for nuclei near the neutron drip line.
 \item Fission is important for abundances of nuclei
       with not only $A > 250$ but $60 < A < 180$.
 \item Light {\it p}-nuclei, such as \nuc{Se}{74}, \nuc{Kr}{78},
       \nuc{Sr}{84}, and \nuc{Mo}{92}, 
       are abundantly produced in the ejecta 
       due to successive proton capture.
       Heavy {\it p}-nuclei, \nuc{In}{113}, \nuc{Sn}{115}, and \nuc{La}{138}, 
       are abundantly synthesized in the jets through fission.
 \item The abundances of {\it p}-nuclei in the ejecta are much greater
       than those in core-collapse SNe, in particular, 
       \nuc{Mo}{92}, \nuc{In}{113}, \nuc{Sn}{115}, and \nuc{La}{138}, 
       which are deficient in the SNe, 
       are significantly produced in the ejecta.
 \item If we assumed that the ejecta via the jets is uniformly mixed 
       to the ISM of $4 \times 10^5 \Ms$, which is pre-enriched by 
       a hyparnova of $40\Ms$, 
       $\log \epsilon_i$ of the material in the ISM well reproduces
       those of the extremely metal poor star, CS31082-001,
       for elements $i$ with $Z = 10$-$92$.
\end{enumerate}

\acknowledgements{
This work was supported in part by the Japan Society for
Promotion of Science(JSPS) Research Fellowships (K.K.), 
Grants-in-Aid for the Scientific Research from the Ministry of
Education, Science and Culture of Japan (No.S14102004, No.14079202, 
No.17540267), and Grant-in-Aid for the 21st century COE program
``Holistic Research and Education Center for Physics of Self-organizing
 Systems''.
We are grateful to K. Arai for his carefully reading the manuscript 
and giving useful comments.
}



\begin{thebibliography}{}

\bibitem[Anders \& Grevesse(1989)]{anders89}
Anders, E., \& Grevesse, N. 1989, Geochim. Cosmochim. Acta 53, 197

\bibitem[Argast et al.(2004)]{argast04} 
Argast, D., Samland, M. Thielemann, F.-K., \& Qian, Y.-Z.\ 2004, 
\aap, 416, 997 

\bibitem[Arnould(1976)]{arnould76}
Arnould, M. 1976, A\&A, 46, 117

\bibitem[Arlandini et al.(1999)]{arlandini99} 
Arlandini, C., K{\"a}ppeler, F., Wisshak, K., Gallino, R., Lugaro, M., 
Busso, M., \& Straniero, O.\ 1999, \apj, 525, 886 

\bibitem[Beloborodov(2003)]{beloborodov03}
Beloborodov, A.~M. 2003, \apj, 588, 931

\bibitem[Blinnikov et al.(1996)]{bdn96} 
Blinnikov, S.~I., Dunina-Barkovskaya, N.~V., \& Nadyozhin, D.~K.\ 1996, 
\apjs, 106, 171

\bibitem[Burris et al.(2000)]{burris00} 
Burris, D.~L., Pilachowski, C.~A., Armandroff, T.~E., Sneden, C., 
Cowan, J.~J., \& Roe, H.\ 2000, \apj, 544, 302 

\bibitem[Clayton(1968)]{clayton68}
Clayton, D.~D. 1968, {\itshape Principles of Stellar Evolution and Nucleosynthesis}
(Newyork: MacGraw-Hill).

\bibitem[Di Matteo, Perna, \& Narayan(2002)]{dpn02}
Di Matteo, T., Perna, R., \& Narayan, R.\ 2002, \apj, 579, 706

\bibitem[Firmani et al.(2004)]{firmani04} 
Firmani, C., Avila-Reese, V., Ghisellini, G., \& Tutukov, A.~V.\ 
2004, \apj, 611, 1033 


\bibitem[Freiburghaus et al.(1999)]{frt99} 
Freiburghaus, C., Rosswog, S., \& Thielemann, F.-K.\ 1999, \apjl, 525, L121 
 

\bibitem[Fujimoto et al.(2003a)]{fujimoto03a}
Fujimoto, S., Hashimoto, M., Koike, O., Arai, K., \& Matsuba, R. 2003a, \apj, 585, 418

\bibitem[Fujimoto et al.(2003b)]{fujimoto03b}
Fujimoto, S., Hashimoto, M., Arai, K., \& Matsuba, R. 2003b, 
{\itshape Origin of Matter and Evolution of the Galaxies 2003} 
ed. M. Terasawa et al., 
pp.344-353 (Singapore: World Scientific).

\bibitem[Fujimoto et al.(2004)]{fujimoto04}
 -----------. 2004, \apj, 614, 817

\bibitem[Fujimoto et al.(2005)]{fujimoto05}
 -----------. 2005, Nucl. Phys. A758, 47

\bibitem[Fujimoto et al.(2006)]{fujimoto06}
Fujimoto, S., Kotake, K., Yamada, S., Hashimoto, M., \& Sato, K.\ 2006, 
preprint, Astro-ph/0602457 (ApJ accepted)

\bibitem[Fuller, Fowler, \& Newman(1980)]{ffn80} 
Fuller, G.~M., Fowler, W.~A., \& Newman, M.~J. 1980, \apjs, 42, 447

\bibitem[Fuller, Fowler, \& Newman(1982)]{ffn82} 
 -----------. 1982, \apjs, 48, 279


\bibitem[Galama et al.(1998)]{galama98}
Galama, T., et al. 1998, Nature, 395, 670

\bibitem[Giavalisco et al.(2004)]{giavalisco04} 
Giavalisco, M., et al.\ 2004, \apjl, 600, L103 

\bibitem[Granot \& Ramirez-Ruiz(2004)]{granot04} 
Granot, J., \& Ramirez-Ruiz, E.\ 2004, \apjl, 609, L9 

\bibitem[Hashimoto(1995)]{hashimoto95}
Hashimoto, M. 1995, Prog. Theor. Phys. 94 663.


\bibitem[Heger et al.(2003)]{heger03} 
Heger, A., Fryer, C.~L., Woosley, S.~E., Langer, N., 
\& Hartmann, D.~H.\ 2003, \apj, 591, 288

\bibitem[Hjorth et al.(2003)]{hjorth03}
Hjorth, J., et al. 2003, \nat, 423, 847

\bibitem[Horiguchi, Tachibana, \& Katakura(1996)]{JAERI} 
Horiguchi, T., Tachibana, T., \& Katakura, T. 1996, 
Chart of the Nuclides (Ibaraki: Nucl. Data Center)

\bibitem[Howard et al.(1991)]{hmw91} 
Howard, W.~M., Meyer, B.~S., \& Woosley, S.~E.\ 1991, \apjl, 373, L5 

\bibitem[Hurley, Sari \& Djorgovski(2004)]{hurley04} 
Hurley K., Sari R. \& Djorgovski S.G., 2004, 
to appear in Compact Stellar X-ray Sources,
eds. W.H.G. Lewin, M. van der Klis, Cambridge Univ. Press (astro-ph/0211620)

\bibitem[K$\ddot{\textrm{a}}$ppeler et al.(1989)]{kap89}
K$\ddot{\textrm{a}}$ppeler, F., Beer, H. and Wisshak, K. 1989,
Rep. Prog. Phys. 52, 945

\bibitem[Kinsey et al.(1996)]{Nudat2.0} 
Kinsey, R. R. et al., 
The NUDAT/PCNUDAT Program for Nuclear Data,
the 9th International Symposium of Capture-Gamma-ray Spectroscopy and Related Topics,
Budapest, Hungary, October (1996) 

\bibitem[Kodama \& Takahashi(1975)]{KT1975} 
Kodama, T., $\&$ Takahashi, K. 1975, Nucl. Phys. A239 489.

\bibitem[Kotake et al.(2004)]{kotake04} 
Kotake, K., Sawai, H., Yamada, S., \& Sato, K. 2004, \apj, 608, 391

%

\bibitem[Langanke et al.(2003)]{langanke03} 
Langanke, K., et al.\ 2003, Phys. Rev. Lett., 90, 241102

%


\bibitem[Maeda \& Nomoto(2003)]{maeda03}
Maeda, K., \& Nomoto, K. 2003, \apj, 598, 1163


\bibitem[Mamdouh, Pearson, Rayet \& Tondeur(1999)]{Mamdouh1999}
Mamdouh, A., Pearson, J.M., Rayet, M.,  \& Tondeur, F. 1999, \nphysa. 644, 389

\bibitem[Mamdouh, Pearson, Rayet \& Tondeur(2001)]{Mamdouh2001}
Mamdouh, A., Pearson, J.M., Rayet, M.,  \& Tondeur, F. 2001, \nphysa. 679, 337


\bibitem[Mizuno, Yamada, Koide, \& Shibata(2004a)]{mizuno04a}
Mizuno, Y., Yamada, S., Koide, S., \& Shibata, K. 2004a, \apj, 606, 395

\bibitem[Mizuno, Yamada, Koide, \& Shibata(2004b)]{mizuno04b}
Mizuno, Y., Yamada, S., Koide, S., \& Shibata, K. 2004b, \apj, 615, 389

\bibitem[Murakami et al.(2005)]{murakami05} 
Murakami, T., Yonetoku, D., Umemura, M., Matsubayashi, T., 
\& Yamazaki, R.\ 2005, \apjl, 625, L13

\bibitem[Nagataki et al.(1997)]{nagataki97}
Nagataki, S., Hashimoto, M., Sato, K., \& Yamada, S. 1997, \apj, 486, 1026

\bibitem[Nishimura et al.(2005)]{nishimura05} 
Nishimura, S., Kotake, K., Hashimoto, M., Yamada, S., Nishimura, N., 
Fujimoto, S., \& Sato, K.\ 2005, 
preprint, Astro-ph/0504100 (ApJ submitted)


\bibitem[Panov et. al(2005)]{panov05}
Panov, I.V., Kolbe, E., Pfeiffer, B. Rauscher,T., Kratz,K.-L.,
\& Thielemann, F.-K.\ 2005, Nucl.Phys. A747, 633

\bibitem[Paczy{\'n}sky \& Wiita(1980)]{pw80}
Paczy{\'n}sky, B., \& Wiita, P.~J. 1980, \aap, 88, 23

\bibitem[Podsiadlowski et al.(2004)]{podsiadlowski04} 
Podsiadlowski, P., Mazzali, P.~A., Nomoto, K., Lazzati, D., 
\& Cappellaro, E.\ 2004, \apjl, 607, L17 

\bibitem[Proga et al.(2003)]{proga03}
Proga, D., MacFadyen, A.~I., Armitage, P. J., \& Begelman, M.,C. 2003, 
\apjl, 599, 5

\bibitem[Pruet, Woosley \& Hoffman(2003)]{PWH03}
Pruet, J., Woosley, S.~E., \& Hoffman, R.~D. 2003a, \apj, 586, 1254

\bibitem[Pruet, Thompson \& Hoffman(2004)]{pruet04}
Pruet, J., Thompson, T.~A., \& Hoffman, R.~D. 2004, \apj, 606, 1006

\bibitem[Qian(2003)]{qian03}
Qian, S. 2003, Prog. Part. Nucl. Phys. 50, 153

\bibitem[Qian \& Wasserburg(2003)]{qw03} 
Qian, Y.-Z. \& Wasserburg, G. J. 2003, \apj, 588, 1099


\bibitem[Rayet et al.(1995)]{rayet95}
Rayet, M., Arnould, M., Hashimoto, M., Prantzos, N., \& Nomoto, K.
1995, A\&A, 298, 517


\bibitem[Shen et al.(1998)]{shen98}
Shen, H., Toki, H., Oyamatsu, K., \& Sumiyoshi, K. 1998, Nucl. Phys. A., 637, 435

\bibitem[Shigeyama \& Tsujimoto(1998)]{shigeyama98} 
Shigeyama, T., \& Tsujimoto, T.\ 1998, \apjl, 507, L135 

\bibitem[Stone \& Norman(1992)]{sn92}
Stone, J. M., \& Norman, M. L. 1992, \apjs, 80, 791

\bibitem[Stone \& Pringle(2001)]{sp01} 
Stone, J.~M., \& Pringle, J.~E.\ 2001, \mnras, 322, 461 

\bibitem[Staudt \& Klapdor-Kleingrothaus(1992)]{Staudt1992} 
Staudt, A., Klapdor-Kleingrothaus, H. V. 1992, Nucl. Phys. A549 254

\bibitem[Sumiyoshi et al.(2000)]{sumi00} 
Sumiyoshi, K., Suzuki, H., Otsuki, K., Terasawa, M., 
\& Yamada, S.\ 2000, \pasj, 52, 601 
 

\bibitem[Sumiyoshi et al.(2001)]{sumi01} 
Sumiyoshi, K., Terasawa, M., Mathews, G.~J., Kajino, T., Yamada, S., 
\& Suzuki, H.\ 2001, \apj, 562, 880 


\bibitem[Terasawa et al.(2002)]{tera02} 
Terasawa, M., Sumiyoshi, K., Yamada, S., Suzuki, H., 
\& Kajino, T.\ 2002, \apjl, 578, L137 
 
\bibitem[Totani(2003)]{totani03} 
Totani, T.\ 2003, \apj, 598, 1151 

\bibitem[Truran et al. (2002)]{truran02} 
Truran, J. W., Cowan, J. J.,Pilachowski, C. A., \& Sneden, C. 2002, 
\pasp, 114, 1293

\bibitem[Yonetoku et al.(2004)]{yonetoku04} 
Yonetoku, D., Murakami, T., Nakamura, T., Yamazaki, R., Inoue, A.~K., 
\& Ioka, K.\ 2004, \apj, 609, 935 

\bibitem[Wanajo et al.(2003)]{wanajo03} 
Wanajo, S., Tamamura, M., Itoh, N., Nomoto, K., Ishimaru, Y., 
Beers, T.~C., \& Nozawa, S.\ 2003, \apj, 593, 968

\bibitem[Wasserburg et al.(1996)]{wass96} 
Wasserburg, G.~J., Busso, M., \& Gallino, R.\ 1996, \apjl, 466, L109 


\bibitem[Wheeler et al.(1998)]{wheeler98} 
Wheeler, J.~C., Cowan, J.~J., \& Hillebrandt, W.\ 1998, \apjl, 493, L101

\bibitem[Woosley \& Howard(1978)]{wh78}
Woosley, S.~E., \& Howard, W.~M., 1978, ApJS, 36, 285

\bibitem[Woosley(1993)]{w93}
Woosley, S. E. 1993, \apj, 405, 273

\bibitem[Woosley \& Weaver(1995)]{ww95}
Woosley, S. E., \& Weaver, T. A. 1995, \apjs, 101, 181

\bibitem[Zeh, Klose, \& Hartmann (2004)]{zkh04}
Zeh, A., Klose, S., \& Hartmann, D. H. 2004, \apj, 609, 952


\end{thebibliography}
\end{document}